\documentclass[%
 reprint,
showpacs,
 amsmath,amssymb,
 aps,
pra,
floatfix,
]{revtex4-1}

\usepackage{graphicx}
\usepackage{dcolumn}
\usepackage{bm}
\usepackage{hyperref}

\usepackage{xspace}
\usepackage{tikz}

\usepackage{qcircuit}

\newcommand{\ER}{Erd\"os-R\'enyi }
\DeclareMathOperator*{\argmin}{arg\,min}
\newcommand{\GW}{Goemans-Williamson\xspace}
\newcommand{\MAXCUT}{{\sc MaxCut}\xspace}	

\newcommand{\Gate}[1]{\textsf{\small #1}}

\begin{document}

\title{\texorpdfstring{Performance of the\\ Quantum Approximate Optimization Algorithm\\ on the Maximum Cut Problem}{Performance of the Quantum Approximate Optimization Algorithm on the Maximum Cut Problem}}

\author{Gavin E. Crooks}
\email{gavin@rigetti.com}
\affiliation{Rigetti Computing, 775 Heinz Avenue, Berkeley, CA 94710, USA}

\date{\today}

\begin{abstract}
The Quantum Approximate Optimization Algorithm (QAOA) is a promising approach for programming a near-term gate-based hybrid quantum computer to find good approximate solutions of hard combinatorial problems.  
However, little is currently know about the capabilities of QAOA, or of the difficulty of the requisite parameters optimization. 
Here, we study the performance of QAOA on the \MAXCUT combinatorial optimization problem, optimizing the quantum circuits on a classical computer using automatic differentiation and stochastic gradient descent, using \texttt{QuantumFlow}, a quantum circuit simulator implemented with \texttt{TensorFlow}.
We find that we can amortize the training cost by optimizing  on batches of problems instances; that QAOA can exceed the performance of the classical polynomial time \GW algorithm with modest circuit depth, and that performance with fixed circuit depth is insensitive to problem size. Moreover, \MAXCUT QAOA can be efficiently implemented on a gate-based quantum computer with limited qubit connectivity, using a qubit swap network.
These observations support the prospects that QAOA will be an effective method for solving interesting problems on  near-term quantum computers.
 \end{abstract}




\maketitle

\paragraph*{Introduction --}

The development of quantum computers continues apace, with the prospect of that intermediate scale machines that exhibit a clear advantage over classical computers will arrive presently~\cite{Preskill2018a}. However, our understanding of how to develop noise resilient algorithms that can run on near-term quantum resources remains limited.
For a few computational problems, we know that a gate-based quantum computer can out-perform any conceivable classical algorithm~\cite{Nielsen2000a}. But for most problems we do not know how to implement effective algorithms on noisy, near term architectures with a modest number of qubits.

One promising approach to solving combinatorial optimization problems on near-term machines is the Quantum Approximate Optimization Algorithm (QAOA)~\cite{Farhi2014a,Farhi2014b,Farhi2016a,Lin2016a,Wecker2016a,Farhi2017a,Guerreschi2017a,Hadfield2017a,Jiang2017a,Yang2017a,Verdon2017a,Zahedinejad2017a,Otterbach2017a,Wang2018a,Dalzell2018a,Lechner2018a,Dumitrescu2018a,Ho2018a,Anschuetz2018a,Fingerhuth2018a}. 
This is a heuristic method, which can be thought of as a time-discretization of adiabatic quantum computing~\cite{Albash2018a}. 
Like a number of near-term approaches to quantum computing~\cite{Peruzzo2014a, McClean2016a, Kandala2017a,Kim2017a}, QAOA is a hybrid classical-quantum algorithm that combines quantum circuits, and classical optimization of those circuits.  The objective is a functional of the quantum state, which in turn is a function parameterized by one and two-qubit gates whose character can be continuously varied.
We perform classical optimization on these continuous gate parameters to generate distributions with significant support on the optimal solution.  The use of a quantum resource allows us to prepare a distribution over an exponentially large sample space in a short sequence of gates, and the use of classical optimization expands the range of problems that we can explore, and may lead to circuits that are relatively robust to imperfections in the implementation of the quantum computer.  
Rather than attempt to program the quantum computer, we instead train the computer to perform the task at hand.

However, because QAOA is a heuristic algorithm it is difficult to provide general complexity theoretic guarantees about the performance of QAOA compared to classical algorithms~\cite{Farhi2014a,Farhi2014b,Lin2016a,Jiang2017a};  of the number of gates needed for effective implementations; nor of the difficulty of the requisite optimization step. In this paper, we use classical simulation with automatic differentiation~\cite{Tamayo-Mendoza2017a} of parameters and stochastic gradient descent to explore these issues. 

\paragraph*{Maximum cut --}
In this paper we study the performance of QAOA on the maximum cut (\MAXCUT) combinatorial optimization problem: Given a graph $G=(V,E)$  with nodes $V$ and edges $E$, find a subset  $S \in V$ such that the number of edges between $S$ and $S\setminus V$ is maximized. 
This problem can be reduced to that of finding the ground state of an antiferromagnetic Ising model.
\MAXCUT is in the APX-complete complexity class: Finding an exact solution is NP-hard~\cite{Karp1972a}, but there are efficient polynomial time classical algorithms that find an approximate answer within some fixed multiplicative factor of the optimum~\cite{Papadimitriou1991a}.
 For \MAXCUT the polynomial time Goemans-Williamson algorithm guarantees an approximation ratio of 0.8785~\cite{Goemans1995a}, which is optimal assuming the unique games conjecture~\cite{Khot2007a}. It is NP-hard to approximate \MAXCUT better than $16/17\approx0.9412$~\cite{Arora1998a, Hastad2001a}.

\paragraph*{QAOA \MAXCUT~\cite{Farhi2014a, Farhi2016a, Guerreschi2017a, Hadfield2017a, Farhi2017a, Otterbach2017a,Wang2018a,Dalzell2018a} --}
To implement \MAXCUT on a quantum computer using QAOA, we encode the graph structure into a cost Hamiltonian which is diagonal in the computational basis, and for which any bit string gives an energy which is the negative of the number of cut edges. 
\begin{equation}
H_C = \frac{1}{2} \sum_{i, j \in E} C_{ij} (1- \sigma^z_i \sigma^z_j)
\label{cost}
\end{equation} 
Here $\sigma^z_i$ is the Pauli Z matrix $(\begin{smallmatrix}1 & 0 \\ 0 & \text{-}1\end{smallmatrix})$ applied to qubit~$i$, $E$ is the set of edges, and  $C$ is the adjacency matrix of the graph, with $C_{ij}=1$  if nodes are connected, and zero otherwise. (We can generalize this discussion to the problem of \MAXCUT on weighted graphs by substituting an edge weight matrix.)  

We prepare the quantum computer in the state with uniform superposition of bit strings by applying a Hadamard gate to each qubit in the zero state. 
Then, for each of $P$ steps of the QAOA algorithm, we evolve the system with the cost Hamiltonian for some angle $\gamma_p$, 
$
U_p = \exp({-i\gamma_p H_C})$, 
 and then evolve the system with a driver Hamiltonian
\begin{equation}
H_{\text{D}} = \frac{1}{2} \sum_i \sigma^x_i
\label{drive}
\end{equation}
 for an angle $\beta_p$, 
$
V_p = \exp({-i\beta_p H_D})
$, 
where $\sigma^x_i$ is the Pauli X matrix $(\begin{smallmatrix}0 & 1 \\ 1 & 0\end{smallmatrix})$.

\begin{figure}[t] 
   \centering
   \includegraphics[width=3.25in]{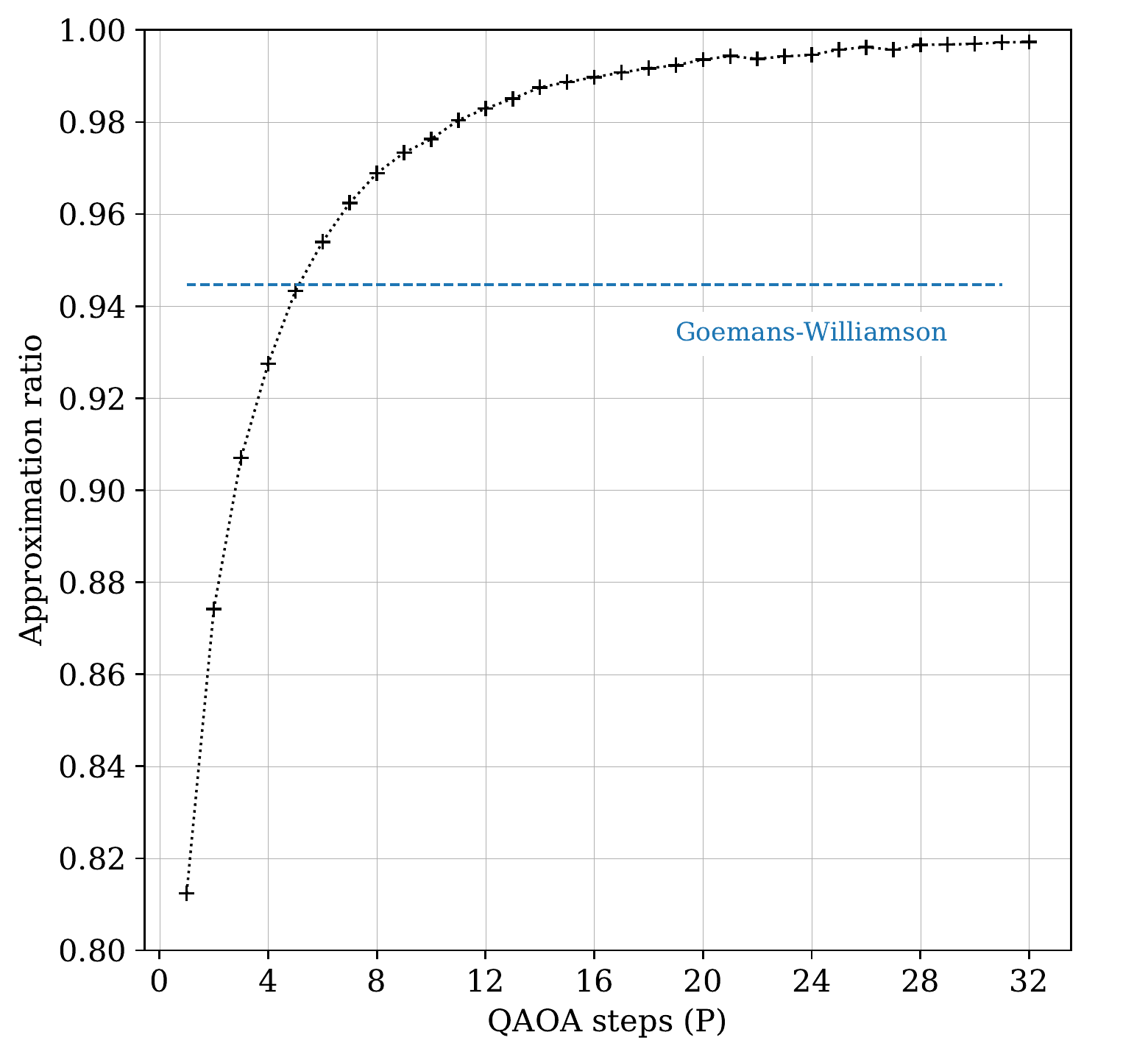} 
	\caption{
	The  average approximation ratio of QAOA on the \MAXCUT optimization problem for 10 node graphs, as a function of QAOA steps ($P$). 
	Each point represents an independently optimized protocol obtained via stochastic gradient descent.
  	The training data consists of 100 graphs from the \ER ensemble with edge probability 50\%, and the test data an independently sampled collection of 100 graphs from the same ensemble.
	For this problem size, QAOA with 5 steps matches the average performance of the classical, polynomial time Goemans-Williamson algorithm on the same data set.
	}
   \label{fig:ARvP}
\end{figure}

\begin{figure}[t] 
   \centering
   \includegraphics[width=3.25in]{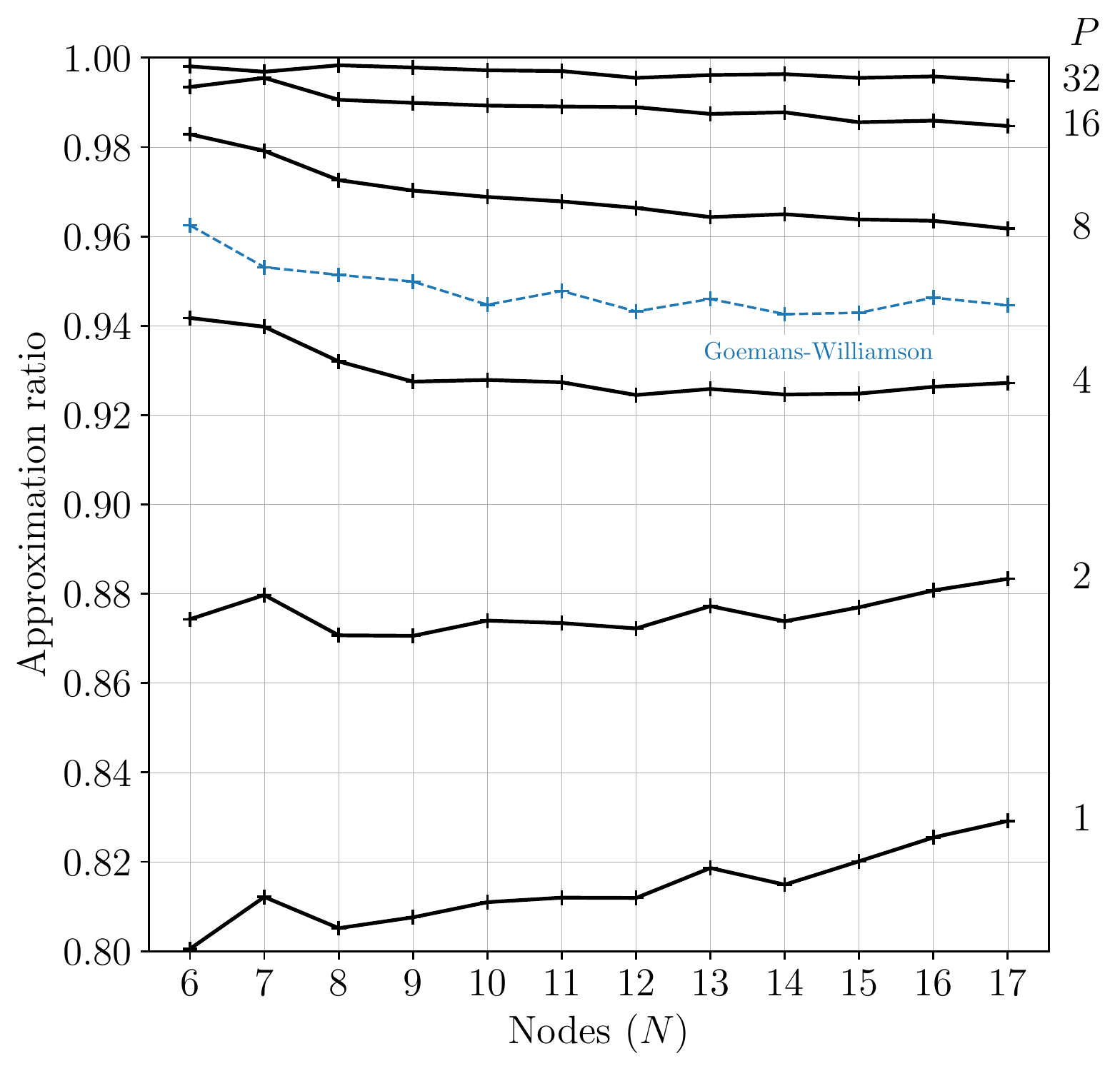} 
   \caption{
   The approximation ratio of QAOA on the \MAXCUT optimization problem, as a function of graph size ($N$) and QAOA steps ($P$). 
   We also show the performance of the classical, polynomial time Goemans-Williamson algorithm on the same test sets. 
   The performance of QAOA increases with circuit depth, and significantly exceeds that of Goemans-Williamson by $P=8$. 
   The performance of QAOA degrades as the graphs get bigger. 
   However, quantum circuits with $P\geq8$ maintain their relative performance advantage over the classical algorithm.
   }
   \label{fig:ARvN}
\end{figure}

Repeated applications of the cost and driver  dynamics evolves the system to the quantum state
\begin{equation}
|\bm{\beta},\bm{\gamma}\rangle = V_P U_P \cdots V_2 U_2 V_1 U_1 |\psi\rangle \ .
\end{equation}

 Functional evaluation is performed by Monte-Carlo estimation of the cost Hamiltonian with samples drawn from the distribution given by $\rho(\bm{\beta}, \bm{\gamma}) = \vert \bm{\beta}, \bm{\gamma}\rangle \langle \bm{\beta}, \bm{\gamma} \vert$.  Each quantum measurement provides access to a random variable $h_{C}^{i}$ which is the cost Hamiltonian evaluated with the computational basis eigenstate resulting from a quantum measurement $\vert i \rangle$.  To obtain an $\epsilon^{2}$ variance on the estimator, $\widehat{\langle  H_C \rangle}_{\rho(\bm{\beta}, \bm{\gamma})}$, one can naively bound the number of samples, which are generated by a quantum state preparation and measurement, by applying the central limit theorem.  Recall that the \MAXCUT problem can be encoded in a linear combination of $2$-local Hamiltonians
\begin{align}
H = \sum_{m=1}^{M}h_{m}P_{m}
\end{align}
where $P_{m}$ is a tensor product of $Z$ operators acting on at most $2$-qubits and $h_{m}$ is the strength of the interaction.  

In the original QAOA algorithm the control parameters $(\bm{\beta},\bm{\gamma})$ are optimized such that the functional, the expectation of the cost Hamiltonian for a given instance of the problem, $\langle \bm{\beta},\bm{\gamma}| H_C |\bm{\beta},\bm{\gamma}\rangle$, is minimized. 
However, this introduces a potentially expensive training step for every query. In the alternative, we train on batches of graphs drawn from the same statistical ensemble, and find a protocol $(\bm{\beta}^\text{opt},\bm{\gamma}^\text{opt})$ that is effective for an entire class of problem instances,
\begin{equation}
(\bm{\beta}^\text{opt},\bm{\gamma}^\text{opt}) =
-\argmin_{\bm{\beta},\bm{\gamma}}
 \frac{1}{|\mathcal{T}|} \sum_{C\in\mathcal{T}} \langle \bm{\beta},\bm{\gamma}| H_{\text{C}} |\bm{\beta},\bm{\gamma}\rangle 
 \ .
\end{equation}
Here $\mathcal{T}$ in the training set of graph adjacency matrices, and $|\mathcal{T}|$ is the number of graphs in the training set. 
A similar approach was used by Wecker {\it et.\ al}~\cite{Wecker2016a}.

We draw the initial QAOA parameters from a normal distribution. If the distribution of these initial parameters is overly broad, then the quantum circuits would be essentially random  and  hard to train, since the dynamics would be chaotic and information could not propagate through the network~\cite{McClean2018a}. Similar issues occurs with deep neural networks~\cite{Glorot2010a,Goodfellow2016a}. 
Empirically, we find that a standard deviation of $0.01$~\cite{Hinton2012a} and a mean of 0.5 (to avoid a symmetry about zero in the parameter space~\cite{Wang2018a}) appears satisfactory.

\paragraph*{Training --}
In order to explore quantum circuits for variational quantum algorithms, we implemented a simple quantum virtual machine (a simulation of a gate based quantum computer) on top of \texttt{TensorFlow}~\cite{Abadi2016a, QF}. This modern, high performance tensor processing library allows us to perform automatic differentiation of the performance metric with respect to the parameters of the quantum circuit. We can therefore train our quantum circuits using back-propagation and stochastic gradient descent.  Code implementing our algorithm is available online~\cite{QF-QAOA}.

Stochastic gradient descent has proved to be extremely effective in machine learning for training deep neural networks~\cite{Goodfellow2016a}. This makes SGD an attractive option for efficient exploration of quantum circuits using classical simulation due to accelerated optimization and ease of use~\cite{Sels2018a}. Ideally, gradients would be calculated with respect to the entire dataset, but this is generally computationally expensive. Instead, at each step of the optimization we calculate the gradient with respect some random subsample of the full dataset.

However, it is difficult to implement stochastic gradient descent directly on a quantum computer, since the requisite gradients are expensive to measure~\cite{Guerreschi2017a},
requiring many observations for each gradient component.
A variety of approaches have been used to optimize QAOA circuits, including Nelder-Mead~\cite{Guerreschi2017a,Verdon2017a}, Monte-Carlo~\cite{Wecker2016a, Yang2017a}, quasi-Newton~\cite{Guerreschi2017a,Anschuetz2018a}, gradient descent~\cite{Wang2018a}, and Bayesian~\cite{Otterbach2017a} methods. It remains an important open question as to the most 
effective and efficient approach to training variational quantum algorithms on a quantum computer.

The validation sets for each graph size consists of 100  randomly generated  graphs draw from the \ER ensemble (Edge probability 50\%). 
Training was performed on an independently sampled collection of 100 graphs drawn from the same distribution. 
There is some chance of overlap between, and redundancy within, the test and validation sets. However, this will only be a significant issue for the smallest graphs, since the number of unique graphs grows rapidly with size, e.g.\ there are over $10^6$ unique 10 node graphs. 
We use the Adam variant of stochastic gradient descent (which includes  momentum and adaptive learning rates)~\cite{Kingma2014a, Goodfellow2016a}, with a mini-batch size of 1, a step size of 0.01, and other parameters at default. 
Both  the $\gamma $ and $\beta$ parameters are periodic~\cite{Farhi2014a}, but we do not constrain these values during optimization. Training from random initialization typically requires at most a few tens of epochs.

\begin{figure}[t] 
   \centering
   \includegraphics[width=3.25in]{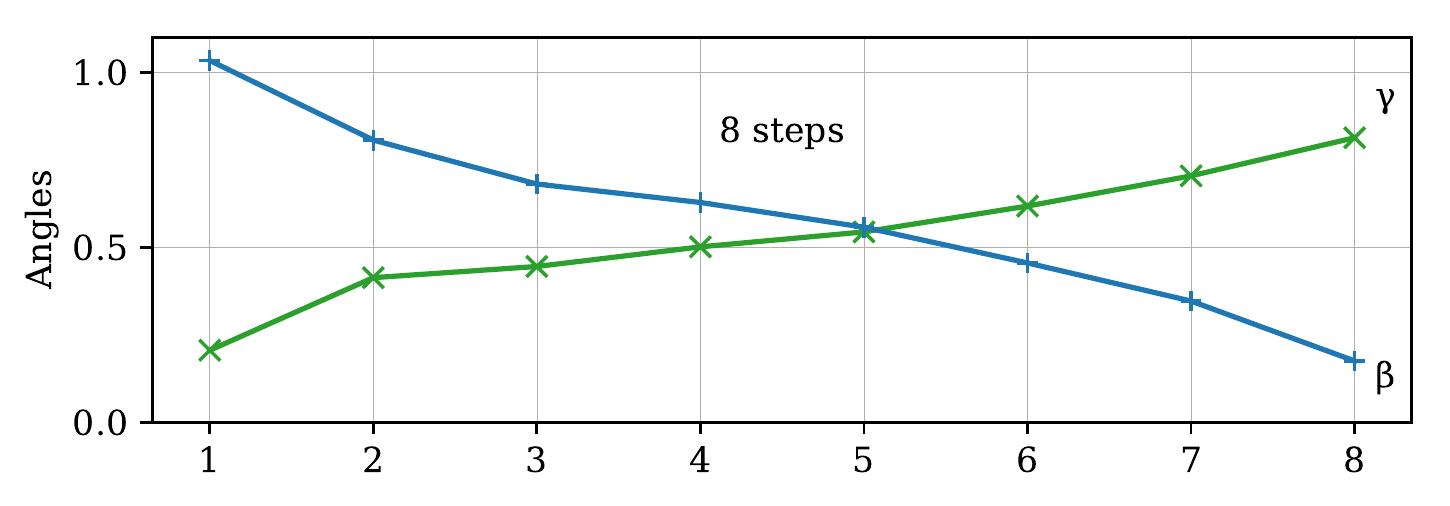} 
   \includegraphics[width=3.25in]{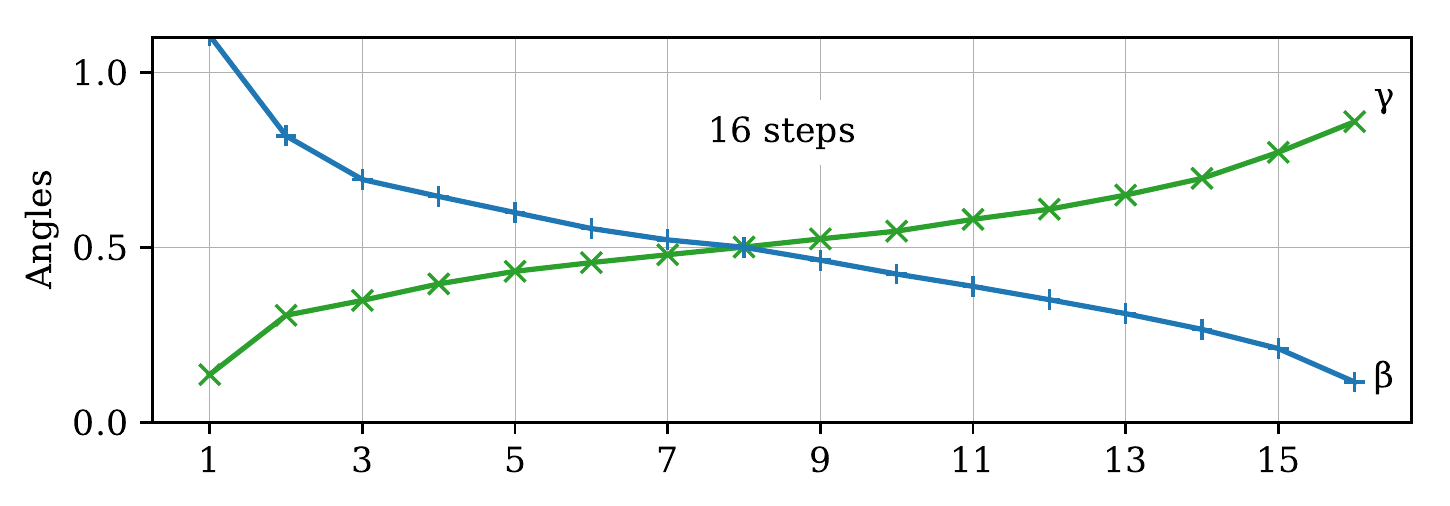}  
   \includegraphics[width=3.25in]{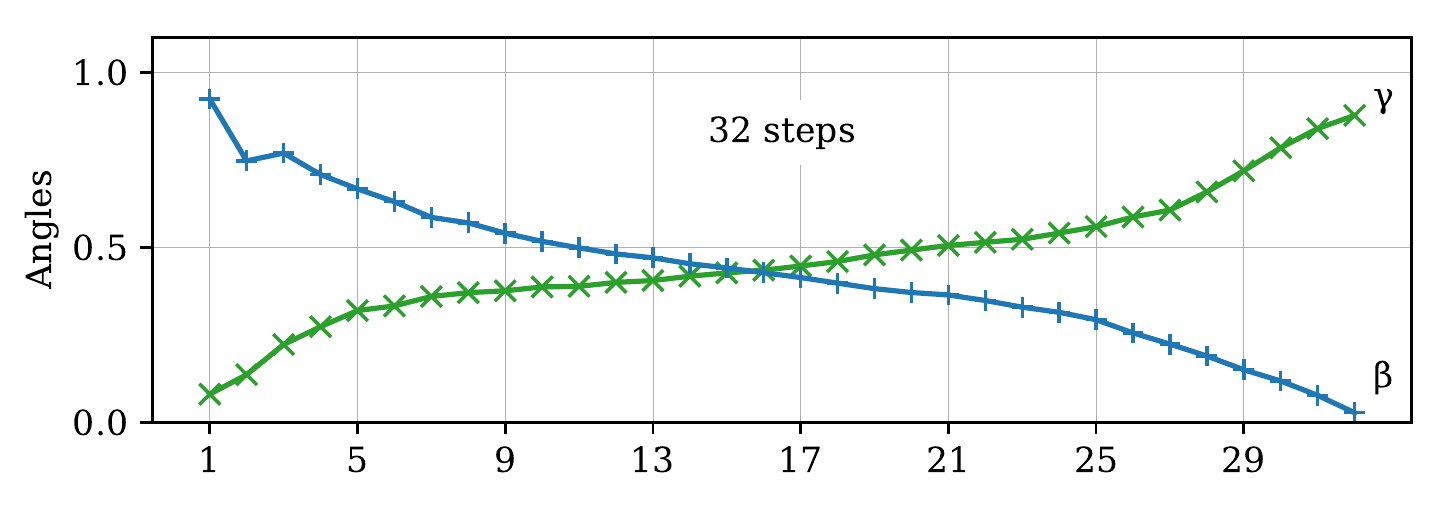}   
   \caption{
 Examples of optimized protocols $(\bm{\beta}^{\text{opt}},\bm{\gamma}^{\text{opt}})$ for \MAXCUT QAOA  on 10 node graphs, with 8, 16, and 32 QAOA steps. The resultant protocols are similar to a linear annealing schedule~\cite{Albash2018a}.
}  
  \label{fig:protocols}
\end{figure}

\paragraph*{Performance --}
In Fig.~\ref{fig:ARvN} we show the approximation ratio as a function of graph size ($N$) and QAOA steps~($P$). Also shown is the average performance of the classical, polynomial time Goemans-Williamson algorithm on the same test sets. 
The performance of QAOA increases with circuit depth, and significantly exceeds that of Goemans-Williamson by $P=8$. 
Although the approximation ratio does vary with problems size, circuits with $P\geq8$ maintain their relative performance advantage over Goemans-Williamson.  Notably there is no indication of a strong dependence of performance on the graph size for the \ER ensemble. 
Typical optimal protocols for different QAOA steps are illustrated in Fig.~\ref{fig:protocols}.

\paragraph*{Computational resources --}
Since we can amortize the training cost, an analysis of the computation complexity of QAOA can focus on the number of gates needed to implement a single instance of the quantum evolution. 
And since the required number of QAOA steps ($P$) for a given performance does not appear to be strongly dependent on the problem size, the limiting resource is the number of two-qubit gates needed to implement a single round of the algorithm.  
We focus on the number of 2-qubit gates, since we require at most two generic 1-qubit rotations per two-qubit gate.

Each of the Pauli-X interactions in the driver Hamiltonian~\eqref{drive} can be implemented with a single one-qubit gate,
\begin{equation}
e^{-\frac{i}{2}\beta \sigma_i^x} \equiv  \begin{array}{c}
\Qcircuit @C=0.5em @R=1.5em {
&\qw &\qw & \gate{{R}_x(\beta)}  & \qw &\qw  
}
\end{array}
\label{X}
\end{equation}
and each of the two-qubit ZZ interactions in the \MAXCUT cost Hamiltonian~\eqref{cost} can be implemented with two \Gate{CNOT} gates, plus a local one-qubit gate.
\begin{equation}
e^{-\frac{i}{2}\gamma(1-\sigma_i^z\sigma^z_j)} \equiv  \begin{array}{c}
\Qcircuit @C=0.5em @R=1.5em {
&\qw &\qw & \ctrl{1} &  \qw  &  \ctrl{1} &\qw & \qw  \\
&\qw &\qw & \targ &  \gate{{R}_z(-\gamma)}  &  \targ & \qw &\qw  
}
\end{array}
\label{ZZ}
\end{equation}
Thus, for a fully connected graph of~$N$ nodes we require $N$ qubits, and $N(N-1)P$ \Gate{CNOT}s.

These resource requirements assume that we can apply gates directly between any two qubits, but in practice qubits are typically arranged in a 2-dimensional lattice with gates only between nearest neighbors. \Gate{SWAP} gates can be used to move logical qubits into proximity, and naively one would expect that every logical gate would require $O(\sqrt{N})$ physical gates~\cite{Cheung2007a}. 
However, we can efficiently implement QAOA with a linear array of qubits using a \Gate{SWAP} network~\cite{Kivlichan2018a,Babbush2018a}, with $O(N)$ overhead.

Suppose we have a linear array of qubits, and we apply \Gate{SWAP} gates between all neighboring pairs in a brickwork pattern.

\begin{center}
\begin{tikzpicture}[scale=0.9]
\tiny 
\node[draw, circle] at (0,0) (q0) {1};
\node[draw, circle] at (0,-1) (q1) {6};
\node[draw, circle] at (0,-2) (q2) {2};
\node[draw, circle] at (0,-3) (q3) {5};
\node[draw, circle] at (0,-4) (q4) {3};
\node[draw, circle] at (0,-5) (q5) {4};

\node[draw, circle] at (5,0) (f0) {6};
\node[draw, circle] at (5,-1) (f1) {5};
\node[draw, circle] at (5,-2) (f2) {1};
\node[draw, circle] at (5,-3) (f3) {4};
\node[draw, circle] at (5,-4) (f4) {2};
\node[draw, circle] at (5,-5) (f5) {3};

\draw (q0) -- (f0);
\draw (q1) -- (f1);
\draw (q2) -- (f2);
\draw (q3) -- (f3);
\draw (q4) -- (f4);
\draw (q5) -- (f5);

\draw[->] (q0.south east)  to [out=-45, in=45] (q2.north east) ;
\draw[->] (q2.south east)  to [out=-45, in=45] (q4.north east) ;
\draw[->] (q4.south east)  to [out=-45, in=45] (q5.north east) ;

\draw[->] (q5.north west)  to [out=-225, in=225] (q3.south west) ;
\draw[->] (q3.north west)  to [out=-225, in=225] (q1.south west) ;
\draw[->] (q1.north west)  to [out=-225, in=225] (q0.south west) ;

\draw (2,0) -- (2,-1);
\draw (2,-2) -- (2,-3);
\draw (2,-4) -- (2,-5);
\draw (3,-1) -- (3,-2);
\draw (3,-3) -- (3,-4);

\node at (3.1,-5.5) (swap) {\textsf{SWAP} gate};
\draw[->] (swap) -- (2.1, -4.5) ;
\node at (5,-5.75) (qubit) {qubits};
\draw[->] (qubit) -- (f5.south) ;

\node at (2.5, 0.5) {Cyclic shift operation};

\normalsize
\node at  (2,0) {$\times$};
\node at  (2,-1) {$\times$};
\node at  (2,-2) {$\times$};
\node at  (2,-3) {$\times$};
\node at  (2,-4) {$\times$};
\node at  (2,-5) {$\times$};
\node at  (3,-1) {$\times$};
\node at  (3,-2) {$\times$};
\node at  (3,-3) {$\times$};
\node at  (3,-4) {$\times$};
\end{tikzpicture}
\end{center}

\noindent This gate architecture implements a ring buffer of qubits. Each application of the \Gate{SWAP} network rotates the qubits one position in the buffer. A full cycle of the buffer requires $N$ applications of this cyclic shift operation, for a total of $N(N-1)$ \Gate{SWAP} gates.

The power of this architecture comes from noting that after half a cycle every qubit has been exchanged with every other qubit. 
At that point in the circuit, when any given pair of qubits are neighbors, we can also interleave the ZZ interactions required by the QAOA algorithm. This is possible because all of the ZZ interactions within a single QAOA iteration commute with one another, so the order that they are applied is unimportant. Each interaction requires 2 \Gate{CNOT}s [Eq.~\eqref{ZZ}], and a \Gate{SWAP} gate can be implemented with 3 \Gate{CNOT} gates. But we can combine the ZZ and \Gate{SWAP} gate into a 2-qubit parametric swap gate (\Gate{PSWAP})~\cite{Smith2016a}. And the \Gate{PSWAP} gate, as with any 2-qubit gate, can also be implemented with at most three \Gate{CNOT} gates~\cite{Vatan2004a, Vidal2004a, Zhang2004a, Shende2004a}. Therefore the QAOA MAXCUT algorithm requires only $\tfrac{3}{2}N(N-1)P$ \Gate{CNOT} gates  with a linear array of qubits. This is only a 50\% overhead compared to the fully connected qubit topology.

\paragraph*{Conclusions --}

Given that we can amortize the training cost; that we can exceed classical performance with a rather modest number of steps; and that QAOA can be efficiently implemented despite limited qubit connectivity,
we expect QAOA \MAXCUT executed on a gate based quantum computer will require $O(N^2P)$ gates, and have a run time of $O(N P)$ (assuming we can apply $O(N)$ gates in parallel). At least for the small \ER graphs studied here, the requisite scaling of QAOA steps~$P$ with node number $N$ appears to be sublinear.  In contrast, \GW requires a running time of $\tilde{O}(Nm)$ for irregular graphs (ignoring logarithmic factors), where $m$ is the number of edges~\cite{Haribara2016a}.

Clearly, these observations are suggestive only, since it is prohibitively expensive to classically simulate the quantum \MAXCUT algorithm on anything but small graphs. 
A fair evaluation against state-of-the-art classical heuristic algorithms, such as simulated annealing and coherent Ising Machines~\cite{Hamerly2018a,Haribara2016a}, would require graph problems with hundreds or thousands of nodes.
Nonetheless, we are optimistic that QAOA can provide a significant quantum advantage~\cite{Farhi2016a,Dalzell2018a} on combinatorial optimization problems with modest numbers of qubits and modest gate depth. Definitive proof will have to await the anticipated arrival of a supremacy class quantum computer with sufficient gate fidelity~\cite{Preskill2012a,Reagor2018a}.

\paragraph*{Acknowledgements --}
We heartily thank 
Matthew Harrigan,
Jonathan Ward,
Keri McKiernan,
Chris Wilson,
and
Marcus da Silva
for astute discussions.
The circuit simulation framework of \texttt{QuantumFlow} is based upon Nicholas Rubin's \texttt{reference-qvm}~\footnote{\url{https://github.com/rigetticomputing/reference-qvm}}, who also developed the \GW benchmark and contributed to the circuit depth analysis.

\bibliography{extra,on_qaoa}

\end{document}